\documentclass[english,pra]{revtex4}
\usepackage[T1]{fontenc}
\usepackage[latin9]{inputenc}
\usepackage{amsmath}
\usepackage{amssymb}

\makeatletter

\usepackage{babel}

\begin{document}

\title{Unambiguous Examining the Orthogonality of Multiple Quantum States}

\author{Shengshi Pang and Shengjun Wu}

\affiliation{Hefei National Laboratory for Physical Sciences at Microscale and
Department of Modern Physics, University of Science and Technology
of China, Anhui 230026, China}
\begin{abstract}
In this article, we study an opposite problem of universal quantum state comparison,
that is unambiguous determining whether multiple unknown quantum states from a Hilbert space
are orthogonal or not. We show that no unambiguous quantum measurement can
accomplish this task with a non-zero probability. Moreover, we extend
the problem to a more general case, that is to compare how orthogonal
multiple unknown quantum states are with a threshold, and it turns
out that given any threshold this extended task is also impossible
by any unambiguous quantum measurement except for a trivial case.
It will be seen that the impossibility revealed in our problem is
stronger than that in the universal quantum state comparison problem and distinct from those in the existing {}``no-go'' theorems.
\end{abstract}


\maketitle
The quantum information science has developed rapidly in the recent
decades, and many fantastic protocols like quantum state teleportation
\cite{teleportation}, quantum dense coding \cite{dense coding},
Shor's factoring algorithm \cite{shor}, etc. rooting in the quantum
superposition principle and quantum entanglement \cite{entanglement}
have demonstrated more excellent abilities of quantum protocols in
communication and computation than classical protocols. However, besides
the advantages of quantum protocols in the information field, the
limitations on quantum information tasks imposed by quantum principles
have also arisen great interest since some {}``no-go'' theorems
were discovered. Among many {}``no-go'' theorems are the well-known
quantum no cloning theorem \cite{nocloning} and quantum no deletion
\cite{nodeleting} theorem. The quantum no cloning theorem states
that an arbitrary unknown quantum state cannot be cloned perfectly
in a deterministic way and it is actually equivalent to that an unknown
quantum state cannot be determined deterministically; and the quantum
no deletion theorem states that an unknown quantum state cannot be
generally deleted and transformed into a {}``blank'' state. These
{}``no-go'' theorems are due to the linearity and the superposition
principle of quantum mechanics and have helped people to have a deeper
insight into the quantum world.

In the classical world, it is generally feasible to compare two or
more arbitrary objects and tell whether they identical or different,
however, Stephen M. Barnett, et al. showed that in the quantum world
it is impossible to compare two arbitrary unknown quantum states and
determine whether they are the same in a perfect or unambiguous way
\cite{orginal pla}. This is another important distinction between
the classical world and the quantum world. After that, the problem
of comparing multiple unknown quantum states has received more intensive
research \cite{jmo,generalization not universal,unambiguous jpa,ensembles 08}
and has been realized in experiment \cite{experiment}.

Enlightened by the universal quantum state comparison problem, we are going
to study a new relevant problem, that is examining whether or not
multiple unknown quantum states from a Hilbert space are orthogonal. This problem is opposite to the universal quantum state comparison problem, and
one may expect a similar but different result as in the universal quantum state comparison
problem. In fact, in studying this problem, we find an impossibility of achieving this task (Theorem 1) and we shall give a short analysis on how this result is compared with that of the universal quantum state comparison problem (Remark 1). Moreover, we shall extend our problem to a more general case, that is comparing
how orthogonal multiple unknown quantum states are with a threshold, and a more complete result will be derived in Theorem 2 and Corollary 1. In the conclusion paragraph, we shall comment on how the impossibility discovered in our problem is different from most existing {}``no-go'' theorems.

We first define our problem in more detail. Suppose there are $n$
arbitrary unknown quantum states $|\psi_{1}\rangle,\cdots,|\psi_{n}\rangle$
from a Hilbert space $\mathcal{H}$ of finite dimension $d$ $\left(d\geq2\right)$,
then we want to study if one can determine whether the $n$ states
are orthogonal or not by quantum measurements. In our research, we
find it always impossible to complete this task by an unambiguous
quantum measurement. {}``Unambiguous'' means that the measurement
may produce an inclusive result with non-zero probability, but if
a result is conclusive it must be error-free. In addition, we shall
assume that the number of the states is no larger than the dimension
of the Hilbert space, i.e., $n\leq d$. This assumption is reasonable,
because if $n>d$ there are no $n$ orthogonal states in the Hilbert
space $\mathcal{H}$ and thus the problem turns to be senseless.

At first glance, the impossibility of achieving the above task seems
obvious. It is known that any unknown quantum state cannot be discriminated
perfectly or unambiguously considering the states in the whole Hilbert
space $\mathcal{H}$ are not linearly independent \cite{discrimination1,discrimination2,discrmimination3},
thus it seems natural that determining whether $n$ arbitrary states
are orthogonal is impossible. However, it should be noted that in
our problem, it actually only requires to obtain the relation between
the $n$ states (i.e., whether the $n$ states are mutually orthogonal)
but not to determine each of the unknown states, so in this point
of view, the impossibility of examining the orthogonality of the $n$
states is beyond the knowledge of the existing {}``no-go'' theorems.

We shall use the \emph{positive operator-valued measure} (POVM) \cite{nielsen}
to study the problem in our research. POVM provides a compact way
to describe a general physical process (no matter the process is local
or non-local) if only the statistical properties of the process are
concerned in the problem. A POVM consists of a set of POVM elements,
each of which is a positive operator and represents a possible outcome
of the physical process, and the POVM elements must sum up to the
identity operator due to the conservation of probability. There are
three possible outcomes of the measurement in our problem: i) the
$n$ unknown states are orthogonal; ii) the $n$ unknown states are
not orthogonal; and iii) the outcome gives no conclusive information.
We shall $R_{1}$, $R_{2}$ and $R_{?}$ to denote the three outcomes
respectively and $M_{1}$, $M_{2}$, $M_{?}$ to denote the corresponding
POVM elements. Note that simultaneous measurement on the whole $n$
states is allowed in our problem, so each $M_{i}$ acts on the composite
Hilbert space $\mathcal{H}^{\otimes n}$ of the $n$ quantum states.
Quantum mechanics tells that the probability that the outcome $R_{i}$
occurs is \begin{equation}
\mathrm{Prob}(R_{i})=\langle\psi_{1}|\otimes\cdots\otimes\langle\psi_{n}|M_{i}|\psi_{1}\rangle\otimes\cdots\otimes|\psi_{n}\rangle.\label{eq:2}\end{equation}
In addition, due to the unambiguity of the measurement, it is required
that $\mathrm{Prob}(R_{1})=0$ when the unknown quantum states are
not orthogonal and $\mathrm{Prob}(R_{2})=0$ when the unknown states
are orthogonal.

The following lemma will be useful to prove the theorems later.

\emph{Lemma 1.} Suppose $\left\{ |e_{1}\rangle,\cdots,|e_{d}\rangle\right\} $
is a basis of a Hilbert space of finite dimension $d$ (but $|e_{1}\rangle,\cdots,|e_{d}\rangle$
are not necessarily orthogonal) and $M$ is a positive semidefinite
operator on that space, if $\langle e_{i}|M|e_{i}\rangle=0$ for all
$i=1,\cdots,d$, then $M=0$.

\emph{Proof.} As the operator $M$ is positive, $M$ can be decomposed
as $M=K^{\dagger}K$, so

\[
\langle e_{i}|M|e_{i}\rangle=\langle e_{i}|K^{\dagger}K|e_{i}\rangle=\left\Vert K|e_{i}\rangle\right\Vert ^{2}=0\]
which implies that\begin{equation}
K|e_{i}\rangle=0,\label{eq:8}\end{equation}
thus\begin{equation}
M|e_{i}\rangle=K^{\dagger}K|e_{i}\rangle=0,\quad i=1,\cdots,d.\label{eq:14}\end{equation}
Considering that $\left\{ |e_{1}\rangle,\cdots,|e_{d}\rangle\right\} $
is a basis, it can be inferred that $M=0$. $\blacksquare$

Now we want to present our first result: determining whether or not
$n$ arbitrary unknown quantum states is impossible with an unambiguous
measurement.

\emph{Theorem 1.} Unambiguous determining whether or not $n$ arbitrary
unknown quantum states are mutually orthogonal in the Hilbert space
$\mathcal{H}$ is impossible.

Theorem 1 actually contains two conclusions:

i) no unambiguous measurement can produce a conclusive result when
the unknown states are orthogonal;

ii) no unambiguous measurement can produce a conclusive result when
the unknown states are not orthogonal.

In the following, we shall prove these two conclusions separately
since the methods to prove them are quite different. We firstly prove
the conclusion i).

\emph{Proof of conclusion i) of Theorem 1.} Let us divide all product
states in the composite Hilbert space $\mathcal{H}^{\otimes n}$ into
two sets: one set contains all product states whose $n$ factor states
are orthogonal, and the other set contains all the remaining product
states in $\mathcal{H}^{\otimes n}$. We denote the first set by $\mathcal{S}_{O}$
and the second by $\mathcal{S}_{N}$. Then, the task of determining
whether $n$ arbitrary quantum states are orthogonal is equivalent
to distinguishing between the sets $\mathcal{S}_{O}$ and $\mathcal{S}_{N}$.
Unambiguity of the measurement requires that when the states are not
orthogonal, the probability that the outcome $R_{1}$ occurs is zero,
so if we can show that the states in $\mathcal{S}_{N}$ span the whole
composite Hilbert space $\mathcal{H}^{\otimes n}$, the first conclusion
can be proven then according to Lemma 1, and this is the core of the
proof.

Now, let us arbitrarily select $n$ quantum states $|\psi_{1}\rangle,\cdots,|\psi_{n}\rangle$
which are not mutually orthogonal, then $|\psi_{1}\rangle\otimes\cdots\otimes|\psi_{n}\rangle\in\mathcal{S}_{N}$.
Suppose $|\phi_{i,j}\rangle$, $j=1,\cdots,d-1$, are $d-1$ orthonormal
states in the orthogonal complement to $|\psi_{i}\rangle$ in $\mathcal{H}$,
then let\begin{equation}
|\psi_{i,j}\rangle=C_{i,j}\left(|\psi_{i}\rangle+\epsilon_{i,j}|\phi_{i,j}\rangle\right),\: j=1,\cdots,d-1,\label{eq:6}\end{equation}
where $\epsilon_{i,j}>0$ and $C_{i,j}=\left(1+\epsilon_{i,j}^{2}\right)^{-\frac{1}{2}}$
is the normalization constant, and denote $|\psi_{i,0}\rangle=|\psi_{i}\rangle$.
It is evident that for each $i=1,\cdots,n$ the $d$ states $|\psi_{i,j}\rangle$,
$j=0,\cdots,d-1$ are linearly independent, and when all $\epsilon_{i,j}$'s
are sufficiently small, the states $|\psi_{1,j_{1}}\rangle,\cdots,|\psi_{n,j_{n}}\rangle$
with arbitrary indexes $j_{1},\cdots,j_{n}=0,\cdots,d-1$ can be still
non-orthogonal, implying $|\psi_{1,j_{1}}\rangle\otimes\cdots\otimes|\psi_{n,j_{n}}\rangle\in\mathcal{S}_{N}$.
Now we choose an arbitrary product state $|\Psi_{1}\rangle\otimes\cdots\otimes|\Psi_{n}\rangle$
from $\mathcal{H}^{\otimes n}$. Since the $d$ states $|\psi_{i,j}\rangle$,
$j=0,\cdots,d-1$ are linearly independent for each $i=1,\cdots,n$,
$|\Psi_{i}\rangle$ can be expanded as $|\Psi_{i}\rangle=\sum_{j_{i}=0}^{d-1}\alpha_{j_{i}}|\psi_{i,,j_{i}}\rangle$
and thus $|\Psi_{1}\rangle\otimes\cdots\otimes|\Psi_{n}\rangle$ has
an expansion of\begin{equation}
|\Psi_{1}\rangle\otimes\cdots\otimes|\Psi_{n}\rangle=\sum_{j_{1},\cdots,j_{n}=0}^{d-1}\alpha_{j_{1}}\cdots\alpha_{j_{n}}|\psi_{1,j_{1}}\rangle,\cdots,|\psi_{n,j_{n}}\rangle.\label{eq:5}\end{equation}
Considering that any state in $\mathcal{H}^{\otimes n}$ can be spanned
by direct product states, the whole direct product space $\mathcal{H}^{\otimes n}$
can be spanned by the states $|\psi_{1,j_{1}}\rangle\otimes\cdots\otimes|\psi_{n,j_{n}}\rangle,\: j_{1},\cdots,j_{n}=0,\cdots,d-1$
according to \eqref{eq:5}, i.e. the $d^{n}$ states $|\psi_{1,j_{1}}\rangle\otimes\cdots\otimes|\psi_{n,j_{n}}\rangle$
forms a basis (not necessarily orthogonal) of $\mathcal{H}^{\otimes n}$.

As $|\psi_{1,j_{1}}\rangle\otimes\cdots\otimes|\psi_{n,j_{n}}\rangle,\: j_{1},\cdots,j_{n}=0,\cdots,d-1,$
are all in the set $\mathcal{S}_{N}$, the unambiguity of the measurement
requires that\begin{equation}
\langle\psi_{1,j_{1}}|\otimes\cdots\otimes\langle\psi_{n,j_{n}}|M_{1}|\psi_{1,j_{1}}\rangle\otimes\cdots\otimes|\psi_{n,j_{n}}\rangle=0,\label{eq:16}\end{equation}
 and thus $M_{1}=0$ according to Lemma 1. This completes the proof
of the first conclusion. $\blacksquare$

Next, we go on to prove the second conclusion of Theorem 1. Like the
proof of the first conclusion, the key to prove the second conclusion
is to show that $\mathcal{S}_{O}$ is a spanning set of the composite
Hilbert space $\mathcal{H}^{\otimes n}$. But note that the method
used in proving the first conclusion will fail to take effect in this
situation because the states $|\psi_{1,j_{1}}\rangle,\cdots,|\psi_{n,j_{n}}\rangle$
constructed in a similar way cannot guarantee to be still orthogonal
no matter how small the $\epsilon_{i,j}$'s are, thus we have to resort
to other methods.

\emph{Lemma 2.} If $n\leq d$, the set $\mathcal{S}_{O}$ can span
the whole composite Hilbert space $\mathcal{H}^{\otimes n}$.

To prove Lemma 2, let us rename the set $\mathcal{S}_{O}$ which contains
all product states whose $n$ factor states are orthogonal as $\mathcal{S}_{O}[\mathcal{H},n]$
to explicitly denote its reliance on $\mathcal{H}$ and $n$, i.e.,

\begin{equation}
\mathcal{S}_{O}[\mathcal{H},n]=\left\{ |\psi_{1}\rangle\otimes\cdots\otimes|\psi_{n}\rangle\mid|\psi_{1}\rangle,\cdots,|\psi_{n}\rangle\in\mathcal{H}\mathrm{\: are\: mutually\: orthogonal}.\right\} .\label{eq:11}\end{equation}
We shall use $\mathcal{V}_{O}[\mathcal{H},n]$ to denote the subspace
spanned by the set $\mathcal{S}_{O}[\mathcal{H},n]$.

Before proving Lemma 2, we need to note the following observation.

\emph{Observation 1.} Let $\Psi_{1,\cdots,n}$ be an arbitrary state
in the subspace $\mathcal{V}_{O}[\mathcal{H},n]$, and $\Psi_{\sigma(1),\cdots,\sigma(n)}$
be the state after an arbitrary permutation $\sigma$ on the $n$
parties of $\Psi_{1,\cdots,n}$, then $\Psi_{\sigma(1),\cdots,\sigma(n)}$
is also in $\mathcal{V}_{O}[\mathcal{H},n]$.

\emph{Proof of Lemma 2.} Suppose $\left\{ |e_{i}\rangle\right\} _{i=1}^{d}$
is an orthonormal basis of $\mathcal{H}$, where $d$ is the dimension
of $\mathcal{H}$. We first prove the lemma for $n=2$. It is equivalent
to prove that all the states $|e_{i}\rangle\otimes|e_{j}\rangle$,
$i,j=1,\cdots,d$, can be expanded by the states in $\mathcal{S}_{O}[\mathcal{H},2]$.
According to the definition of $\mathcal{S}_{O}[\mathcal{H},n]$ \eqref{eq:11},
the states $|e_{i}\rangle\otimes|e_{j}\rangle$, $i\neq j$, belong
to $\mathcal{S}_{O}[\mathcal{H},2]$, so it is sufficient to prove
that the $d$ states $|e_{i}\rangle\otimes|e_{i}\rangle$, $i=1,\cdots,d$,
can be expanded by $\mathcal{S}_{O}[\mathcal{H},2]$. In fact, for
any $|e_{i}\rangle$, $i=1,\cdots,d$, we can always choose an arbitrary
$|e_{j}\rangle$, $j\neq i$, and\begin{eqnarray}
|e_{i}\rangle\otimes|e_{i}\rangle & = & \frac{1}{\sqrt{2}}\left(|e_{i}\rangle+|e_{j}\rangle\right)\otimes\frac{1}{\sqrt{2}}\left(|e_{i}\rangle-|e_{j}\rangle\right)\nonumber \\
 &  & +\frac{1}{\sqrt{2}}\left(|e_{i}\rangle+i|e_{j}\rangle\right)\otimes\frac{1}{\sqrt{2}}\left(|e_{i}\rangle-i|e_{j}\rangle\right)\nonumber \\
 &  & +\frac{1}{2}\left(1+i\right)\left(|e_{i}\rangle|e_{j}\rangle-|e_{j}\rangle|e_{i}\rangle\right).\label{eq:12}\end{eqnarray}
 Therefore, all $|e_{i}\rangle\otimes|e_{i}\rangle$, $i=1,\cdots,d$,
are in the subspace $\mathcal{V}_{O}[\mathcal{H},2]$.

Next, we prove the lemma for $n\geq3$ by induction. We assume that
when $n=2,\cdots,m$ $\left(m\leq d-1\right)$, each product state
$|e_{i_{1}}\rangle\otimes\cdots\otimes|e_{i_{n}}\rangle$ $\left(i_{1},\cdots,i_{n}=1,\cdots,d\right)$
belongs to the subspace $\mathcal{V}_{O}[\mathcal{H}^{\prime},n]$
of $\mathcal{H}^{\otimes n}$ where $\mathcal{H}^{\prime}$ is an
\emph{arbitrary} subspace of dimension $n$ containing all the factor
states $|e_{i_{1}}\rangle,\cdots,|e_{i_{n}}\rangle$. It is explicit
that this assumption is satisfied by $n=2$ and we shall show that
it also holds true for $n=m+1$ below.

For $n=m+1$, it is obvious that when $|e_{i_{1}}\rangle,\cdots,|e_{i_{m+1}}\rangle$
are different from each other, the induction assumption holds, so
we only need to give a proof when at least two of the factor states
$|e_{i_{1}}\rangle,\cdots,|e_{i_{m+1}}\rangle$ are the same. The
product states $|e_{i_{1}}\rangle\otimes\cdots\otimes|e_{i_{m+1}}\rangle$
of which at least two factor states are the same can be further divided
into two categories: one is that the factor states $|e_{i_{1}}\rangle,\cdots,|e_{i_{m+1}}\rangle$
are not all the same, and the other is that the factor states $|e_{i_{1}}\rangle,\cdots,|e_{i_{m+1}}\rangle$
are all the same.

We first prove for the first category. When $|e_{i_{1}}\rangle,\cdots,|e_{i_{m+1}}\rangle$
are not all the same, suppose all distinctive states in $|e_{i_{1}}\rangle,\cdots,|e_{i_{m+1}}\rangle$
are $|e_{i_{1}}\rangle,\cdots,|e_{i_{k}}\rangle$, $2\leq k\leq m$.
Then by performing appropriate permutations on the $m+1$ involved
parties, the state $|e_{i_{1}}\rangle\otimes\cdots\otimes|e_{i_{m+1}}\rangle$
can be re-arranged to\begin{equation}
|e_{i_{1}}\rangle^{\otimes r_{i_{1}}}\otimes\cdots\otimes|e_{i_{k}}\rangle^{\otimes r_{i_{k}}},\label{eq:3}\end{equation}
where $r_{i_{1}}+\cdots+r_{i_{k}}=m+1$ and $1\leq r_{i_{1}},\cdots,r_{i_{k}}\leq m$.
Now let us \emph{arbitrarily} choose an $m+1$-dimensional subspace
$\mathcal{H_{E}}$ of $\mathcal{H}$ that contains the states $|e_{i_{1}}\rangle,\cdots,|e_{i_{k}}\rangle$,
and decompose $\mathcal{H_{E}}$ as\begin{equation}
\mathcal{H_{E}}=\mathcal{H}_{1}\oplus\cdots\oplus\mathcal{H}_{k}\label{eq:13}\end{equation}
 where\begin{equation}
|e_{i_{l}}\rangle\in\mathcal{H}_{l},\:\mathrm{dim}(\mathcal{H}_{l})=r_{i_{l}}\:\mathrm{and}\:\mathcal{H}_{i}\perp\mathcal{H}_{j},\label{eq:7}\end{equation}
 for all $l=1,\cdots,k$ and $i\neq j$. According to the induction
assumption, $|e_{i_{j}}\rangle^{\otimes r_{i_{j}}}\in\mathcal{V}_{O}\left[\mathcal{H}_{j},r_{i_{j}}\right]$,
and since all $\mathcal{H}_{i}$'s are mutually orthogonal,\begin{equation}
|e_{i_{1}}\rangle^{\otimes r_{i_{1}}}\otimes\cdots\otimes|e_{i_{k}}\rangle^{\otimes r_{i_{k}}}\in\bigotimes_{j=1}^{k}\mathcal{V}_{O}\left[\mathcal{H}_{j},r_{i_{j}}\right]\subseteq\mathcal{V}_{O}[\mathcal{H}_{\mathcal{E}},m+1]\label{eq:9}\end{equation}
 thus considering Observation 1 the induction assumption holds for
the product state $|e_{i_{1}}\rangle\otimes\cdots\otimes|e_{i_{m+1}}\rangle$.

Finally, we prove for the states of the second category when $n=m+1$,
that is $|e_{i}\rangle^{\otimes m+1}$, $i=1,\cdots,d$. Let $|e_{j}\rangle$
be an arbitrary state in the basis $\left\{ |e_{i}\rangle\right\} _{i=1}^{d}$
other than $|e_{i}\rangle$, and $\mathcal{H}_{\mathcal{E}^{\prime}}$
be any subspace of dimension $m+1$ containing $|e_{i}\rangle$ and
$|e_{j}\rangle$. Note that\begin{eqnarray}
|e_{i}\rangle^{\otimes m+1} & = & \frac{1}{2}\left(|e_{i}\rangle-|e_{j}\rangle\right)\otimes\left(|e_{i}\rangle+|e_{j}\rangle\right)^{\otimes m}\nonumber \\
 &  & +\frac{1}{2}\left(|e_{i}\rangle-|e_{j}\rangle\right)^{\otimes2}\otimes\left(|e_{i}\rangle+|e_{j}\rangle\right)^{\otimes m-1}\nonumber \\
 &  & -|e_{i}\rangle^{\otimes2}\otimes\sum_{k=1}^{m-1}\sum_{\sigma}\left(|e_{i}\rangle^{\otimes m-k-1}\otimes|e_{j}\rangle^{\otimes k}\right)\nonumber \\
 &  & +|e_{j}\rangle\otimes|e_{i}\rangle\otimes\sum_{k=0}^{m-1}\sum_{\sigma}\left(|e_{i}\rangle^{\otimes m-k-1}\otimes|e_{j}\rangle^{\otimes k}\right),\label{eq:4}\end{eqnarray}
where the sum $\sum_{\sigma}$ runs over all possible permutations
of the $m-1$ factor states in the parentheses excluding the permutations
between the same factor states. Since the terms on the right side
of \eqref{eq:4} are all of the first category, they can be expanded
by $\mathcal{S}_{O}[\mathcal{H}_{\mathcal{E}^{\prime}},m+1]$, so
\begin{equation}
|e_{i}\rangle^{\otimes m+1}\in\mathcal{V}_{O}[\mathcal{H}_{\mathcal{E}^{\prime}},m+1].\label{eq:10}\end{equation}

Thus we complete the induction proof for $n=m+1$, and we can repeat
this step for $m+2,\cdots,d$. Consequently, we can conclude that
the composite space $\mathcal{H}^{\otimes n}$ can be expanded by
$\mathcal{S}_{O}[\mathcal{H},n]$ for all $n=2,\cdots,d$. $\blacksquare$

With Lemma 2 we can prove the second conclusion of Theorem 1 straightforwardly
now.

\emph{Proof of conclusion ii) of Theorem 1.} Suppose $|\psi_{1}\rangle,\cdots,|\psi_{n}\rangle$
are arbitrary orthogonal quantum states in $\mathcal{H}$, the unambiguity
of the measurement requires that\begin{equation}
\mathrm{Prob}(R_{2})=\langle\psi_{1}|\otimes\cdots\otimes\langle\psi_{n}|M_{2}|\psi_{1}\rangle\otimes\cdots\otimes|\psi_{n}\rangle=0.\label{eq:15}\end{equation}
 In another word, Eq. \eqref{eq:15} holds for all product states
in $\mathcal{S}_{O}$.

Since $\mathcal{S}_{O}$ is a spanning set of $\mathcal{H}^{\otimes n}$
by Lemma 2, $M_{2}=0$ as a result of Eq. \eqref{eq:15} and Lemma
1. Therefore, no measurement can unambiguously detect the case of
arbitrary $n$ non-orthogonal states. $\blacksquare$

\emph{Remark 1.} In the universal quantum state comparison problem,
it is known that unambiguous measurement can never give a conclusive
outcome when the states to be compared are identical \cite{orginal pla},
but can probabilistically give a conclusive outcome when the states
are different, and the POVM identifying the case of different states
can be constructed as the projector onto the orthogonal complement
to the totally symmetric subspace in the composite Hilbert space $\mathcal{H}^{\otimes n}$
\cite{unambiguous jpa}. However, in our problem, Theorem 1 shows
that unambiguous measurement cannot give a conclusive outcome not
only when the unknown states are orthogonal but also when they are
not orthogonal, in other words, the task can be never completed in
either case. Therefore, compared with the ordinary universal quantum state comparison
problem, Theorem 1 shows a stronger impossibility in our problem.

In the following, we extend our problem to a more general case as
we have mentioned before, i.e. comparing how orthogonal arbitrary
unknown quantum states are with a given threshold. We first introduce
a natural measure of how orthogonal $n$ states $|\psi_{1}\rangle,\cdots,|\psi_{n}\rangle$
are as the maximum fidelity between the $n$ states

\begin{equation}
\mathrm{max}_{i,j=1,\cdots,n}\left\{ \left|\langle\psi_{i}|\psi_{j}\rangle\right|\right\} \label{eq:0}\end{equation}
and we call it \emph{orthogonality measure}, then we study if it is
possible to determine whether the orthogonality measure of $n$ unknown
states is below or above a given threshold $A$, that is to test whether
the inequality\begin{equation}
\mathrm{max}_{i,j=1,\cdots,n}\left\{ \left|\langle\psi_{i}|\psi_{j}\rangle\right|\right\} \leq A\label{eq:1}\end{equation}
holds or not. We have the following theorem.

\emph{Theorem 2.} Given any threshold $0<A<1$, unambiguous comparison
between the orthogonality measure of arbitrary $n$ unknown quantum
states and the given threshold is impossible.

Theorem 2 also includes two conclusions simultaneously. One is that
no measurement can unambiguously signal that the orthogonal measure
of $n$ unknown states is below the threshold; and the other is that
no measurement can unambiguously signal that the orthogonal measure
of $n$ unknown states is above the threshold. These two conclusions
can be both proved in a similar way as the conclusion i) of Theorem
1, so we skip the details of the proof here and only give an outline
of the proof here.

To prove that no measurement can produce an unambiguous result when
the orthogonal measure of the $n$ states is below the threshold,
one can select such $n$ states $|\psi_{1}\rangle,\cdots,|\psi_{n}\rangle$
from $\mathcal{H}$ that their orthogonal measure is above the threshold,
and then construct $|\psi_{i,j}\rangle=C_{i,j}\left(|\psi_{i}\rangle+\epsilon_{i,j}|\phi_{i,j}\rangle\right),\: j=1,\cdots,d-1$,
$|\psi_{i,0}\rangle=|\psi_{i}\rangle$ with sufficiently small $\epsilon_{i,j}$
so that the orthogonality measure of $|\psi_{1,j_{1}}\rangle,\cdots,|\psi_{n,j_{n}}\rangle$
is still above the threshold for any $j_{1},\cdots,j_{n}=0,\cdots,d-1$.
It can be shown that $|\psi_{1,j_{1}}\rangle\otimes\cdots\otimes|\psi_{n,j_{n}}\rangle$,
$j_{1},\cdots,j_{n}=0,\cdots,d-1$ forms a basis of the composite
Hilbert space $\mathcal{H}^{\otimes n}$ like the proof of the conclusion
i) of Theorem 1, then from the unambiguity of the measurement and
Lemma 1, it can be proved that $M_{1}=0$. And the proof of $M_{2}=0$
is similar.

It should be noted that the case $A=1$ is excluded in Theorem 2 because
when $A=1$ arbitrary $n$ unknown states can always meet the threshold
inequality \eqref{eq:1} and thus a trivial measurement $M_{1}=I^{\otimes n}$,
$M_{2}=M_{3}=0$ can always produce the correct outcome.

Note that we do not include the case $A=0$ in Theorem 2, but in fact
the case $A=0$ is equivalent to Theorem 1, so we have the following
corollary.

\emph{Corollary 1.} Given any threshold $0\leq A<1$, no unambiguous
measurement can give a conclusive result of comparing the orthogonality
measure of arbitrary $n$ unknown states with the given threshold.

In summary, we have studied and proven the impossibility of determining
whether $n$ arbitrary unknown quantum states are mutually orthogonal
or whether the orthogonality measure of $n$ unknown quantum states
$\mathrm{max}_{i,j=1,\cdots,n}\left\{ \left|\langle\psi_{i}|\psi_{j}\rangle\right|\right\} $
meets a given threshold in this article. This problem is opposite
to the universal quantum state comparison and the result of
this problem shows a stronger impossibility than that of the universal
quantum state comparison problem, because in the
universal comparison problem $M_1$ vanishes but $M_2$ exists while
in our problem both $M_1$ and $M_2$ vanish. Moreover, compared
with the existing ``no-go'' theorems, the most distinguishing
feature of this impossibility is that it is a ``collective'' impossibility
involving multiple quantum states simultaneously
and allowing non-local quantum measurements, while most
existing ``no-go'' theorems (e.g., the no-cloning theorem and
the no-deletion theorem) concern one single state each time
and employ only local measurements or operations. We hope our research can help to provide a further
insight into the limitations on quantum information protocols imposed
by quantum principles.

This work is supported by the NNSF of China, the CAS and the National
Fundamental Research Program. S. Pang also acknowledges the support
from the Innovation Foundation of USTC.

\end{document}